\documentclass[conference]{IEEEtran}
\IEEEoverridecommandlockouts
\usepackage{cite}
\usepackage{amsmath,amssymb,amsfonts}
\usepackage{algorithmic}
\usepackage{graphicx}
\usepackage{textcomp}
\usepackage{url}
\usepackage{xcolor}
\usepackage{pifont}
\def\BibTeX{{\rm B\kern-.05em{\sc i\kern-.025em b}\kern-.08em
    T\kern-.1667em\lower.7ex\hbox{E}\kern-.125emX}}
\begin{document}

\title{SecurePay: Enabling Secure and Fast Payment Processing for Platform Economy\\
}

\author{
    \IEEEauthorblockN{Junru Lin\IEEEauthorrefmark{1}, Mingzhe Liu\IEEEauthorrefmark{1}, Songze Li\IEEEauthorrefmark{2}\textsuperscript{‡}, Xuechao Wang\IEEEauthorrefmark{1}\thanks{For correspondence on the paper, please contact Songze Li and Xuechao Wang.}}
    \IEEEauthorblockA{\IEEEauthorrefmark{1}The Hong Kong University of Science and Technology (Guangzhou)}
    \IEEEauthorblockA{\IEEEauthorrefmark{2}School of Cyber Science and Engineering, Southeast University, Nanjing, China}
    \IEEEauthorblockA{\IEEEauthorrefmark{3}Engineering Research Center of Blockchain Application, \\ Supervision and Management (Southeast University), Ministry of Education}
    \IEEEauthorblockA{Email: \{junlin373, mliu007\}@connect.hkust-gz.edu.cn, songzeli@seu.edu.cn, xuechaowang@hkust-gz.edu.cn}
}

\maketitle

\begin{abstract}
Recent years have witnessed a rapid development of platform economy, as it effectively addresses the trust dilemma between untrusted online buyers and merchants. 
However, malicious platforms can misuse users' funds and information, causing severe security concerns.
Previous research efforts aimed at enhancing security in platform payment systems often sacrificed processing performance, while those focusing on processing efficiency struggled to completely prevent fund and information misuse.
In this paper, we introduce SecurePay, a secure, yet performant payment processing system for platform economy.
SecurePay is the first payment system that combines permissioned blockchain with central bank digital currency (CBDC) to ensure fund security, information security, and resistance to collusion by intermediaries; it also facilitates counter-party auditing, closed-loop regulation, and enhances operational efficiency for transaction settlement. We develop a full implementation of the proposed SecurePay system~\cite{cbdc}, and our experiments conducted on personal devices demonstrate a throughput of 256.4 transactions per second and an average latency of 4.29 seconds, demonstrating a comparable processing efficiency with a centralized system, with a significantly improved security level. 
\end{abstract}


\section{Introduction}
The platform economy has grown significantly in popularity among internet users by reducing costs and improving transparency and efficiency in product selection and decision-making.  A key factor in the rise of E-commerce platforms is their ability to mitigate trust issues between buyers and sellers, especially regarding pre-payment and goods delivery. This is achieved through escrow protocols, which position the platform as a trusted intermediary for collecting buyer's payment and distributing them to merchants based on transaction details~\cite{goldfeder}.

\begin{figure}[htbp]
  \centering
  \includegraphics[width=\linewidth]{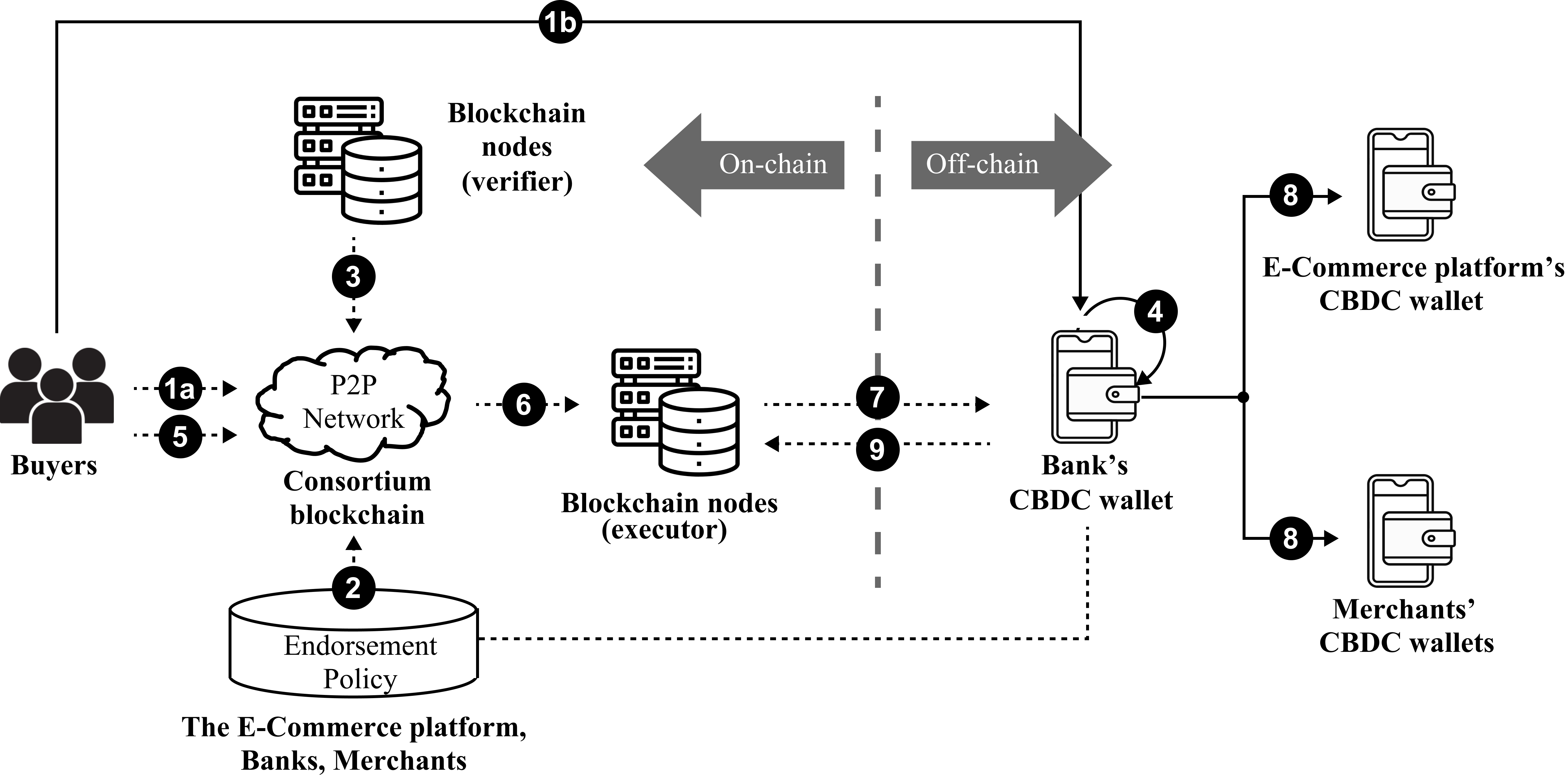}
  \caption{SecurePay system overview. Step 1-8 outlines the key interactions between on-chain and off-chain activities, starting from when a buyer initiates a trade to when the transaction is confirmed and closed.}
  \label{sysover}
\end{figure}

One key issue with the escrow protocol is its reliance on the e-commerce platform to honestly process transaction information and payments for buyers and merchants. This centralized trust model poses significant risks by placing all personal data and transaction details under the control of a single intermediary, inadequately addressing both \textit{fund and information security}.

\textbf{Fund security.} An escrow protocol ensures fund security if it prohibits the mediator from transferring escrowed funds without the consent of either buyer or seller \cite{goldfeder}. In platform-based systems, funds are typically collected and held in the platform's wallet until the buyer confirms receipt of goods. However, the critical period known as escrow time—the duration between fund collection and disbursement to the merchant—poses significant risks. During this period, the platform has unchecked access to the funds~\cite{fy, kim, rail}, which may lead to potential misuse of the funds, undermining the security and trust that the escrow protocol aims to provide \cite{endurthi}.

\textbf{Information security.} An escrow protocol is considered to uphold information integrity when transaction details cannot be altered without the mutual consent of both the buyer and the seller \cite{fy, rail}. However, in E-commerce platforms, where transaction data are stored in centralized databases, this information becomes susceptible to unilateral modifications by the platform operators \cite{kim, endurthi}. Furthermore, the requirement for buyers to provide authenticated information for payment verification on E-commerce platforms may compromise user privacy, as these platforms cannot guarantee the protection of sensitive data against unauthorized access or misuse.

\begin{figure}[htbp]
  \centering
  \includegraphics[width=0.7\linewidth]{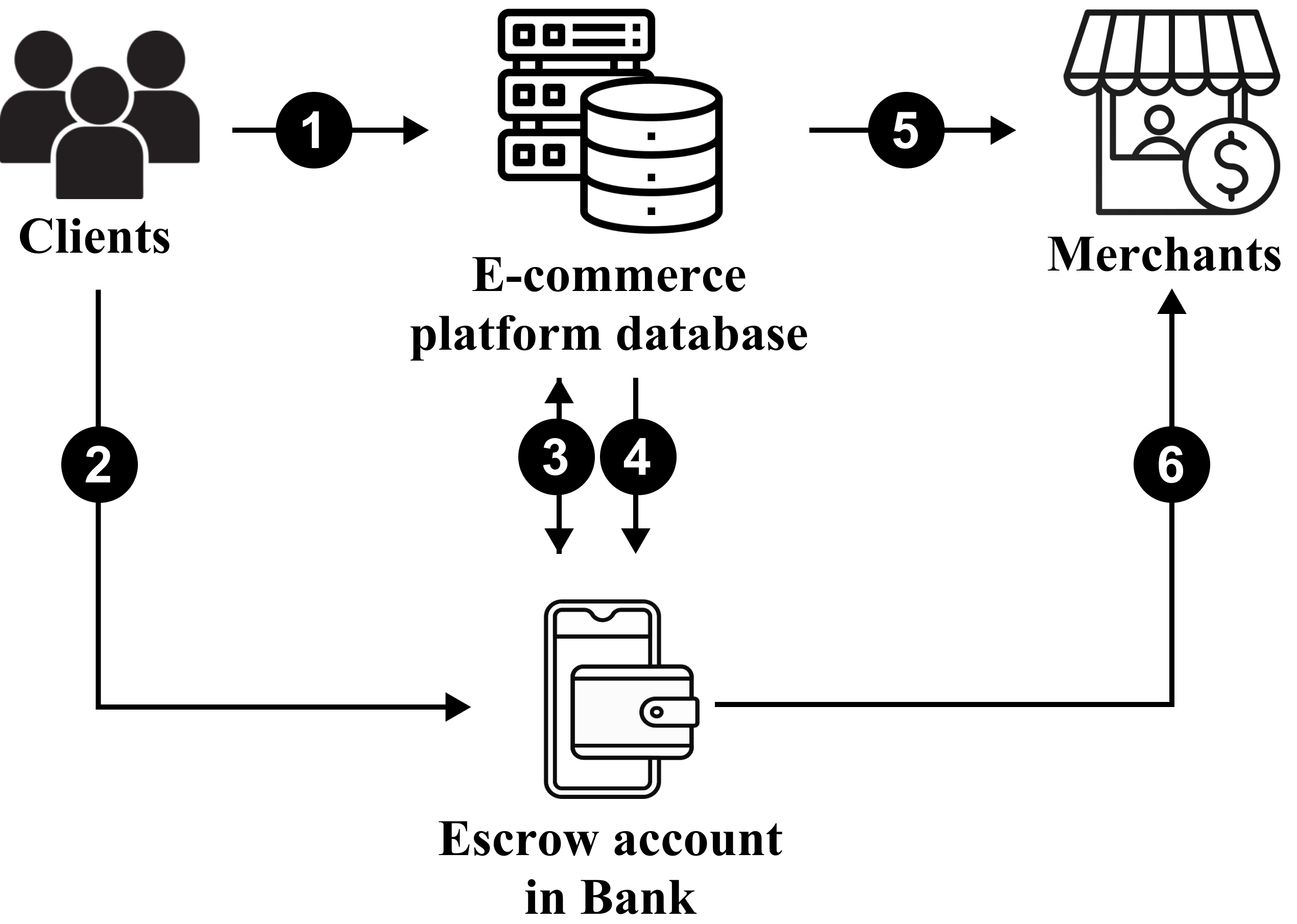}
  \caption{Centralized payment system that relies on regulation for protecting fund security.}
  \label{central}
\end{figure}

\textbf{Proposed solution.} In this paper, we propose \verb|SecurePay|, a payment system designed to provide both fund security and information security for payment processing in the platform economy. To achieve information integrity, we use a permissioned blockchain to delegate information processing to multi-authorized entities, including merchants, banks, and the E-commerce platform. Meanwhile, we harness the programmability and privacy-preserving features of Central Bank Digital Currencies (CBDCs) for managing payments, ensuring fund security and safeguarding user privacy \cite{kumar, bis, jiang, kiayias}. We propose an  on-chain and off-chain joint system (Figure \ref{sysover}) that connects CBDCs with permissioned blockchain.

\textbf{On-chain.} Our on-chain component decentralizes control over information processing and provide transparent visibility to buyers and merchants within the platform economy. We utilize smart contracts for automating escrow transactions and employ a multi-party endorsement protocol to ensure the integrity of transaction information.

\textbf{Off-chain.} The CBDC-based payment system addresses the challenges of fund security and user privacy through its programmability \cite{imf, central, qian}, which locks escrowed funds and automatically releases them upon receiving on-chain confirmation. Additionally, CBDCs offer privacy-preserving mechanisms, enabling the E-commerce platform to validate payments without accessing the buyer's personal information \cite{bis, allen, kiayias}.

The decision to separate payment processing from on-chain activities is driven by two key considerations. Firstly, it helps mitigate the risks associated with using cryptocurrencies that are created and operated on-chain, as these currencies often exhibit high volatility in value \cite{kumar, keister}. Secondly, this separation enhances user privacy by enabling payments to be verified off-chain solely by banks, preventing the exposure of entire wallet balances to unauthorized parties. 

In summary, we make the following contributions:
\begin{itemize}
    \item We propose \verb|SecurePay|, a framework that provides a secure and efficient payment system for E-commerce platforms, ensuring both fund and information security.
    \item We introduce an on-chain and off-chain structure that maintains user data privacy while enabling efficient multi-party verification and collaboration.
    \item We evaluate \verb|SecurePay| at various scales, and experimental results demonstrate that it outperforms previous approaches.
\end{itemize}

\begin{figure}[htbp]
  \centering
  \includegraphics[width=0.7\linewidth]{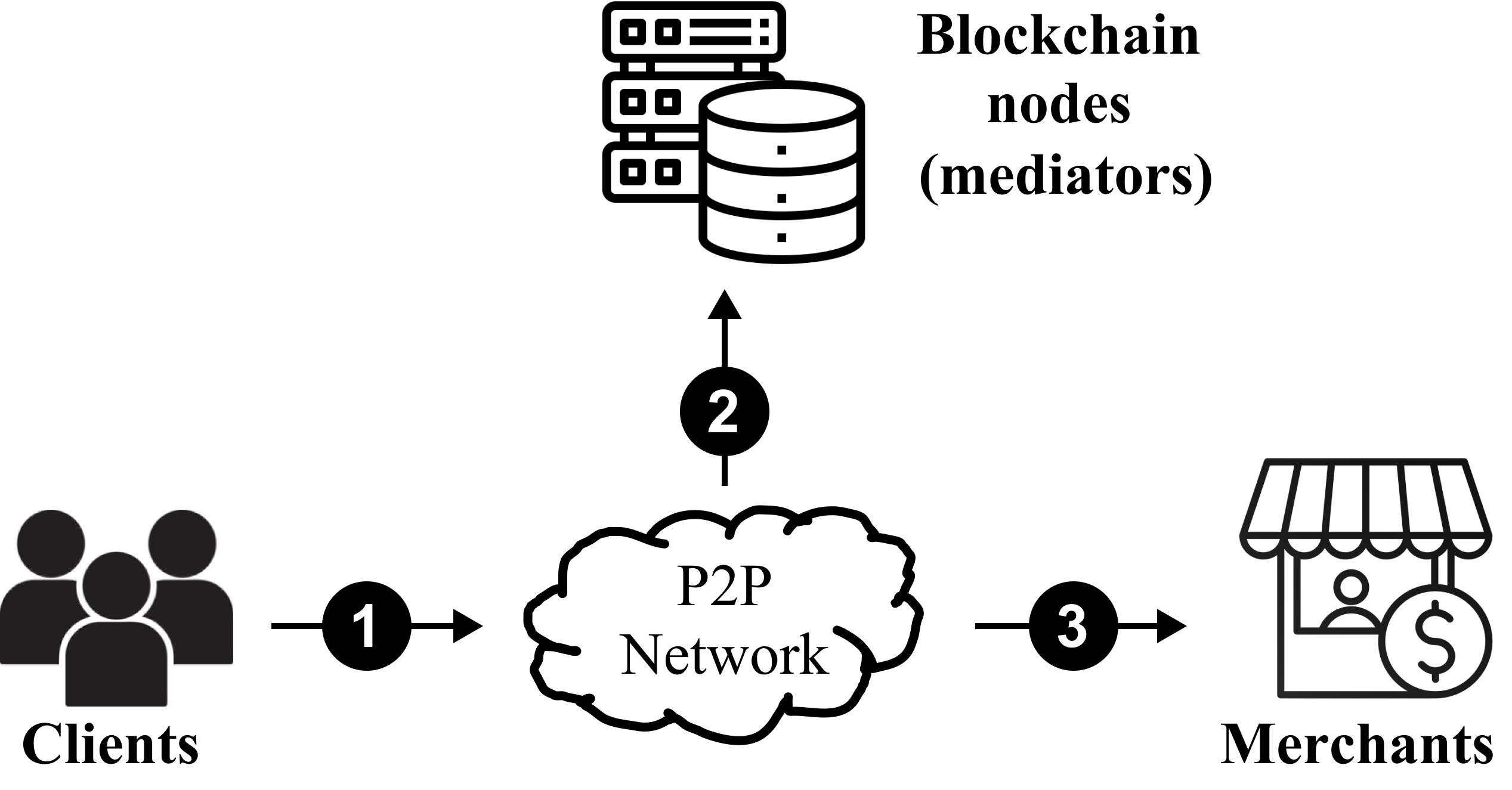}
  \caption{Decentralized payment system that relies on distributed power of recording transaction information.}
  \label{decent}
\end{figure}

\section{Background and motivation}
We explore previous solutions and their limitations within two distinct lines of related work. The first line of solutions focuses on utilizing centralized control and enforcing regula- tions for payment processing. This approach typically involves stringent oversight and management by centralized entities to ensure security and compliance. Conversely, the second line of solutions emphasizes distributing power, using decentralized mediators to replace traditional E-commerce platforms for information and settlement processing. This method leverages distributed ledger technology to enhance transparency and reduce reliance on central authorities.
\subsection{Existing Payment Systems}

\textbf{Centralized-based solutions.} In the domain of centralized payment systems, Fang \textit{et al.} \cite{fy} have proposed a regulated framework that governs how platforms process escrow funds (Figure \ref{central}). Under this framework, the bank establishes an escrow account for the platform to receive and authenticate the buyer's payment (step \ding{172}-\ding{173}). The funds are then held in escrow and only released to the merchant upon confirmation of product shipment (step \ding{175}-\ding{177}).

\textbf{Decentralized-based solutions.} This framework employ a peer-to-peer network for trade recording and utilize cryptocurrency as a medium of exchange for balance clearing, such as Encrypt-and-swap protocol \cite{goldfeder}, and Normachain \cite{liu}. In these systems, transaction information is transmitted directly from buyers to merchants and is subsequently verified by blockchain nodes acting as mediators (Figure \ref{decent} step \ding{172}-\ding{174}). These nodes determine the timing and recipient of the escrow fund release, thereby ensuring transaction integrity without relying on a centralized platform.

\textbf{Limitations of existing solutions.}
we classify existing solutions into two types, including centralized-based solutions and decentralized-based solutions. We elaborate on each type of solution and discuss limitations as follows.

\begin{figure*}
  \includegraphics[width=\textwidth]{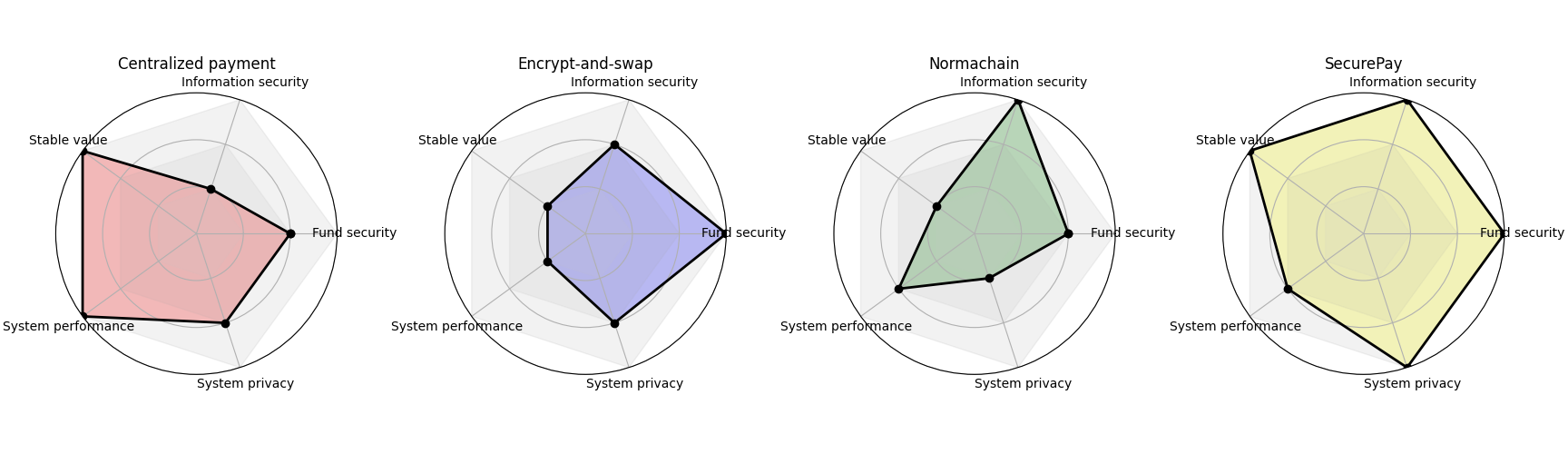}
  \caption{Comparison of previous solutions and SecurePay on five functional indicators (Low-Medium-High).}
  \label{compar}
\end{figure*}

\begin{itemize}
    \item \textbf{Centralized.} These solutions \cite{fy, qian, guo, central, rail} exhibit vulnerabilities in information security due to the management of transaction details by e-commerce platforms, which risks manipulation through unauthorized changes to recipient addresses. Furthermore, sharing transaction information with stakeholders such as banks and courier companies, without sufficient oversight and technical safeguards, exposes user privacy and data security to risks of unauthorized access.
    \item \textbf{Decentralized.} These solutions \cite{liu, goldfeder, kim, joshi, zimbeck, asgaonkar, khan} exhibit vulnerabilities in system performance and stable value of currency transacted. The performance is notably slow due to the time-intensive process of achieving consensus and processing transactions on a distributed network. Additionally, despite encryption efforts, these systems expose user data privacy risks \cite{endurthi} as transaction details are publicly accessible on the blockchain, allowing inferences about transaction amounts and tracing of user wallet balances. 
    
\end{itemize}

\subsection{Central Bank Digital Currency}
Central Bank Digital Currency (CBDC) represents a pivotal advancement in the domain of digital finance, merging the reliability of traditional fiat currencies with the technological innovations characteristic of modern payment systems (summarized in Appendix A) \cite{auer}. Distinct from cryptocurrencies, CBDCs are stabilized through a 1:1 backing by fiat currency, effectively addressing the inherent volatility typically associated with digital assets \cite{kumar, keister}. Furthermore, CBDCs facilitate instant transaction settlements and uphold consumer privacy while adhering to stringent regulatory frameworks, including Anti-Money Laundering (AML)/Combating Financing of Terrorism (CFT) requirements \cite{auer, fy, bis, benigno, imf}. Notably, the paramount feature of CBDCs lies in their programmability via smart contracts \cite{imf}, which enables the creation of escrow-like wallets capable of autonomously locking and releasing funds according to pre-established rules. This capability significantly enhances both the security and efficiency of transactions, positioning programmability as the foundational element for the future of secure digital finance.

\subsection{Privacy-Preserving Techniques in Payment}
Modern solutions protect user data through cryptographic methods and controlled transparency. PUF-Cash \cite{fragkos} uses blind signatures and multi-party trusted authorities to anonymize transaction linkages, preventing identity exposure. ZkLedger \cite{narula} employs zero-knowledge proofs (e.g., zk-SNARKs) to ensure transactional privacy while enabling public auditing. Central Bank Digital Currencies (CBDCs) \cite{bis, benigno, imf, auer} employ cryptographic techniques such as blind signatures and zero-knowledge proofs to enforce controlled visibility of transactional information. This framework ensures that sensitive data remains exclusively accessible to authorized entities (specifically the central bank), while simultaneously permitting verification capabilities for non-authorized third parties such as e-commerce platforms. Crucially, these external auditors can validate payment authenticity and integrity without obtaining access to protected buyer information.

\subsection{Research Motivations}
To address the limitations of existing solutions, it is essential to ensure that e-commerce payment processing systems provide the following properties: fund security, information security, stable value of currency transacted, user data privacy protection, and efficient processing time (Figure \ref{compar}). 

\verb|SecurePay| aims to achieve these properties through the following design features:

\begin{itemize}
    \item Information Security: \verb|SecurePay| introduces multi-party endorsement protocol within a permissioned blockchain to ensure information integrity between users and processing parties.
    \item Fund Security: \verb|SecurePay| introduces a smart contract-based escrow protocol utilizing a CBDC wallet \cite{hamilton} to secure transactions.
    \item System Performance: \verb|SecurePay| develops an off-chain and on-chain structure to separate information processing from payment processing, thereby boosting transaction efficiency.
    \item User's Privacy: \verb|SecurePay| design on-chain and off-chain linkages to guarantee verifiability while maintaining user privacy using CBDC \cite{hamilton}.
    \item Stable Value of Currency Transacted: \verb|SecurePay| utilizes CBDC, which is backed by fiat currency, to provide a stable value for transactions \cite{imf, bis, kumar, keister}.
\end{itemize}

\section{securepay system description}

Figure \ref{sysover} illustrates the architecture and workflow of \verb|SecurePay|. The framework comprises six roles: buyers, blockchain nodes, merchants, couriers, E-commerce platforms, and banks. Their functions are outlined as follows:

\textbf{E-commerce platforms and banks}, as the initiators of the permissioned blockchain, are responsible for creating smart contract templates that facilitate escrow services. These smart contracts are directly linked to off-chain CBDC wallets, which ensures that the payment process is verifiable and transparent. \textbf{Buyers} call and deploy these smart escrow contracts for transaction proposal. Upon deployment, the contracts are published on the network for endorsement by other stakeholders. \textbf{Merchants, E-commerce platforms, and banks} act as endorsers, submitting their digital signatures to validate and authenticate transaction proposals. \textbf{Blockchain nodes} are pivotal in maintaining the security and consistency of the blockchain, ensuring that each proposal has garnered sufficient endorsements and logging contract updates when a smart contract is executed following a client's confirmation of receipt of the product. \textbf{Couriers} are tasked with confirming delivery through their signatures and reach consensus with the buyer on the delivery state on-chain.

\subsection{Key Components}
\begin{figure}[htbp]
  \centering
  \includegraphics[width=\linewidth]{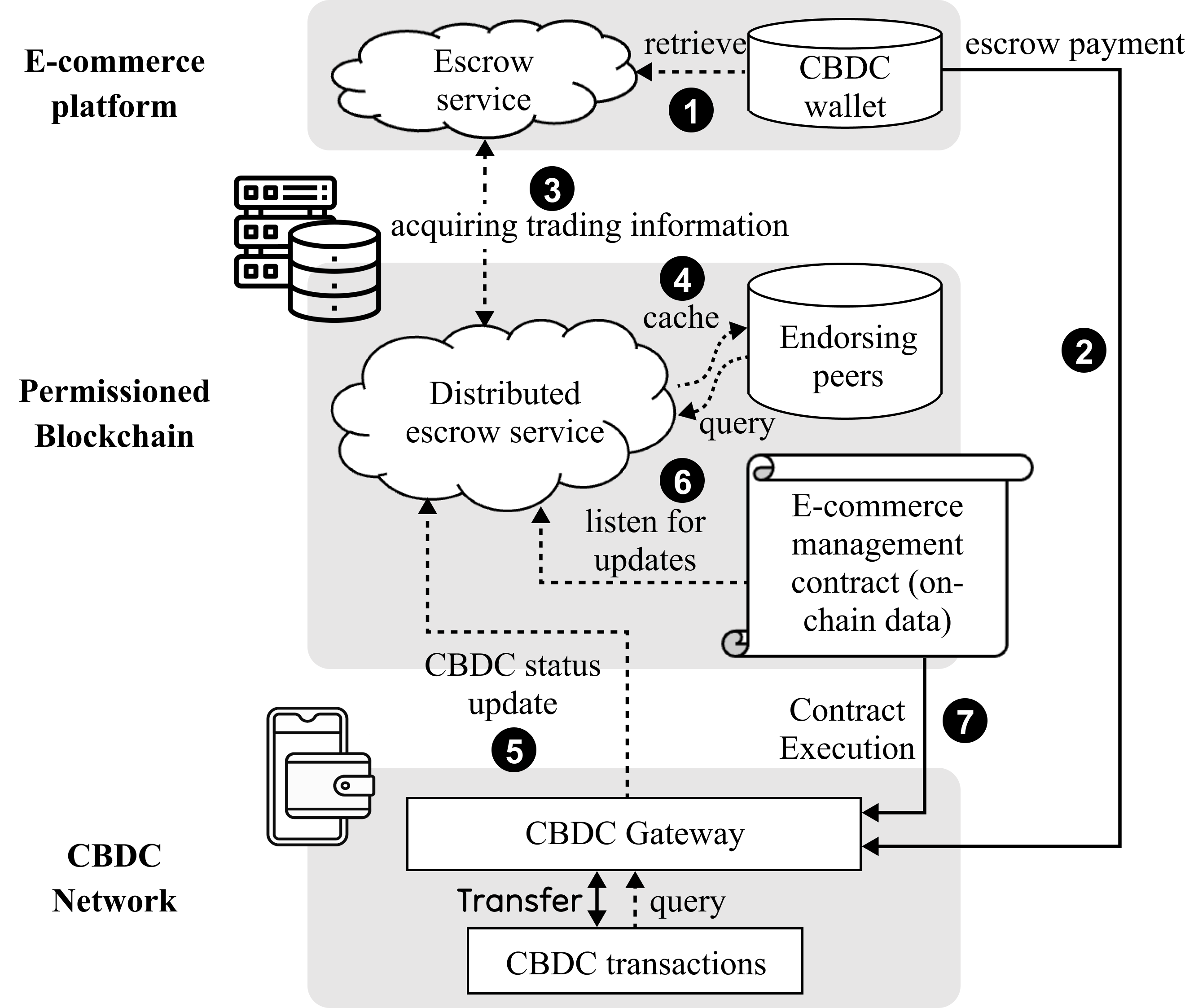}
  \caption{High level technical design of SecurePay. The solid line represents fund processing and the dash line represents information processing.}
  \label{detail}
\end{figure}

The detailed design of our system comprises three key components: the E-commerce platform, the permissioned blockchain, and the CBDC network (Figure \ref{detail}). The interactions between these components are explained as follows:

\textbf{E-commerce Platform.} \label{privacy}
In the proposed E-commerce platform, both buyers, merchants, and couriers are required to register their CBDC wallets via a CBDC gateway, which issues certificates to ensure secure transactions. Buyer's identity will be preserved from linking buyer's ID with CBDC wallet ID, rather than directly linking to user's personal information (step \ding{172}). The CBDC gateway facilitates several key functions: it enables users to \verb|query| their transaction history and check balances, \verb|transfer| funds for purchases, and \verb|redeem| CBDCs into fiat currency (step \ding{173}). Upon the completion of a purchase, the transaction details are compiled by the platform's escrow service and forwarded to a permissioned blockchain network, where it undergoes a multi-party endorsement process.

\textbf{Permissioned Blockchain.}
We use a permissioned blockchain to process transaction information and deploy smart contract to automate escrow service. The version and template of smart contracts are managed through permissioned consensus among key stakeholders such as the banks, and the E-commerce platform. In this setup, buyers can synchronize their trading information from their purchase in E-commerce platform by calling the smart contract as a transaction proposal (step \ding{174}). This proposal is then endorsed by appointed parties including banks, merchants, and the E-commerce platform to ensure \textit{information integrity} (step \ding{175}). The bank verifies the payment in the CBDC network, confirming buyers' payments and endorsing them on the permissioned blockchain without exposing this information to other users on-chain to safeguard \textit{user privacy} (step \ding{176}). The execution of contracts is facilitated by chain's state database, which allows the CBDC network to monitor updates and accordingly settle funds in alignment with on-chain transactions (step \ding{177}).

\textbf{CBDC Network.}
The CBDC network is managed by a consortium of licensed commercial banks that have the authority to handle transactions, interfacing with payment networks, and safeguarding customer funds. This network performs three main functions: Firstly, it stores users' wallet IDs and identity information along with associated certificates (step \ding{173}). Secondly, it establishes a CBDC escrow wallet by leveraging the programmability of CBDCs to automatically lock the escrow payments until the buyer confirms delivery (step \ding{174}). Thirdly, it monitors the contract execution on the permissioned blockchain to facilitate fund transfers on behalf of customers(step \ding{178}).

\subsection{Workflow}

The workflow of the \verb|SecurePay| framework is presented in Fig. \ref{sysover}, we illustrate it as three parts:

\textbf{Transaction proposal.} Buyers submit transaction details, including transaction ID, product ID, product price, and merchants' CBDC wallet address, generated by the E-commerce platform, on-chain via client applications as a transaction proposal (step \ding{172}-a). Upon submission, the transaction information becomes tamper-proof to ensure \textit{information integrity} and is subject to validation through a multi-party endorsement algorithm.

\textbf{Multi-party endorsement.} \label{multi}
The endorsement parties involved in a buyer's transaction proposals consist of three entities: the commercial banks where the buyer and merchant registered their CBDC accounts, the merchants from whom the buyer is making purchases, and the E-commerce platform facilitating the trade. Each E-commerce platform operates on a separate blockchain tailored to its business model and privacy requirements. Buyers broadcast their transaction proposals and request endorsement $REQ_{AUTH}$ accompanied by their digital certificate $CERT_{B_i}$. Upon completing the payment using their CBDC wallet, the buyer's bank confirms the fund transfer, while the merchant's bank simultaneously verifies the presence of funds in the escrow wallet. These banks, along with the E-commerce platform and merchants, endorse the $REQ_{AUTH}$ and broadcast their decision on the blockchain network. The blockchain nodes validate whether all necessary parties have endorsed the proposal for the same result to ensure \textit{information integrity}.

\textbf{Smart contract escrow.} 
If blockchain nodes collect all required parties' endorsements, the smart contract will be deployed. The CBDC escrow wallet is then programmed to be locked to ensure \textit{fund security} (step \ding{175}). When couriers deliver the product, they will submit their digital signature on-chain for the buyer's proposal and forward it to buyer to sign for confirmation (step \ding{176}). The smart contract will execute when buyers confirm delivery (step \ding{177}), prompting the blockchain nodes to update the ledger with contract updates. The locked escrow funds will be released based on the on-chain ledger updates, transferring to the appointed merchant's wallet for their revenue and to the E-commerce platform's wallet for commissions (steps \ding{178}-\ding{179}).

\subsection{Threat Model and Security Goals}\label{threat}
\textbf{Fund security.} We consider a scenario where an attacker operates within the banking system and gains access to the private key of the CBDC escrow wallet. The primary targets in this scenario are the merchants, who are at risk of losing their rightful revenue. The critical time frame for this attack spans the escrow period, extending from the buyer's initial payment to the final settlement to the merchant. During this window, the attacker potentially move the escrow fund to personal accounts without acknowledgement from either the buyer or the merchant. 

\textbf{Information integrity.} We consider attackers originating from within the mediators' systems responsible for processing transaction information. These attackers may attempt to persuade other mediators to collude in tampering with transaction information. These insider threats pose a significant risk to the integrity of the E-commerce ecosystem. The potential victims in this scenario are buyers or merchants who face the risk of financial loss due to malicious transaction data recording. 

\textbf{User privacy.} The primary concern revolves around attackers targeting buyers' personal information, including authenticated identities, payment details, and account balances. Unauthorized access to this sensitive information poses a significant risk of severe online fraud, leading to substantial financial losses for users. We posit that the threat extends to malicious platforms or entities obtaining private user information with the intent to exploit it for personal gain. 

Compared with other blockchain systems, \verb|SecurePay| introduces additional threats due to its unique on-chain and off-chain design, creating new attack vectors that require careful consideration:

\begin{itemize}
    \item Reshipping attack: This type of attack involves key stakeholders such as the buyer and the courier, whose actions are crucial in confirming the delivery of goods. A malicious buyer may engage in a reshipping attack by falsely claiming that the product was not received and requesting the merchant to resend it. 
    \item Settlement Inconsistency: The bank is responsible for processing settlements from the escrow wallet to the merchant's wallet based on the latest contract update on the blockchain. However, if the bank fails to adhere to the latest update, discrepancies between on-chain and off-chain records may occur, potentially leading to revenue loss for the merchant.
\end{itemize}

Considering the above threat model and assumptions, we summarize the security goals of \verb|SecurePay|:

\begin{itemize}
    \item \textbf{Correctness.} \verb|SecurePay| aims to ensure the correctness of transaction processing for platform economy by ensuring fund security and information integrity.
    \item \textbf{Liveness.} The system includes a robust recovery and dispute resolution mechanism that enhances resilience against reshipping attacks and settlement inconsistencies.
    \item \textbf{User Privacy.} \verb|SecurePay| is committed to preventing unauthorized access and misuse of user data, thereby ensuring information security.
\end{itemize}

\section{security analysis}

Based on the threat model proposed in Section \ref{threat}, we analyze the system security formally to show how the aforementioned security goals, i.e., liveness, correctness, and user privacy, are achieved.

\subsection{System Correctness}
In the \verb|SecurePay| framework, the programmability of CBDC and the multi-party endorsement policy play a crucial role in ensuring system correctness. To defend against potential attacks outlined in the threat model, these technologies can be applied in the following ways:

\textbf{Issue 1. Fund security:} The delegation of escrow fund management to banks introduces a vulnerability wherein an internal attacker could unauthorizedly transact the escrow fund during the escrow period.

\textbf{Attack details.} Our threat model assumes the existence of an attacker within the banking system. This individual could gain access to the private key associated with the CBDC escrow wallet. The public availability of the escrow wallet's address on the blockchain, coupled with the attacker's control of the private key, could facilitate the diversion of funds for unauthorized purposes during escrow period. The attacker might then attempt to return the funds upon receiving an on-chain request to release the escrow.

\textbf{Defend strategy.} We leverage the programmable nature of CBDCs to secure the escrow funds against unauthorized withdrawals by intermediaries. The process is as follows: The escrow CBDC will only be released for settlement upon the system's receipt of an on-chain confirmation of delivery from the buyer. Such a mechanism ensures that even in the event an attacker gains possession of the wallet's private key, they will be rendered incapable of misusing the funds in the absence of the required on-chain confirmations.

\textbf{Issue 2. Information Integrity:} Transaction information is maintained through consensus among the merchant, bank, and E-commerce platform. This arrangement introduces a vulnerability that attackers could exploit, persuading collusion to tamper with transaction information.

\textbf{Attack details.} The system exists a potential for collusion aimed at tampering with transaction information before a buyer initiates transactions. An attacker could coax the buyer, platform, and bank into colluding to detrimentally affect the merchant's revenue by initiating fraudulent transaction proposals. Similarly, the attacker could persuade the platform, bank, and merchant to reject a buyer's legitimate transaction proposal.

\textbf{Defend strategy.} Our defense leverages a multi-party endorsement protocol to ensure the accuracy of information, effectively eliminating opportunities for collusion. This requires collecting endorsements from banks, merchants, and E-commerce platforms. In situations where mediators, such as banks and E-commerce platforms, conspire with the buyer against the merchant, the absence of endorsement from the merchant results in the automatic rejection of the transaction proposal. Conversely, should the mediators and the merchant collude to decline a buyer's valid proposal, this proposal will likewise be marked as invalid in the blockchain record. Proposals that manage to gather the required endorsements from all parties involved are recognized as valid by the blockchain nodes, leading to their processing and subsequent smart contract update on-chain. On the other hand, proposals that fail to secure the necessary endorsements trigger the CBDC's refund mechanism, effectively annulling the transactions and eliminating any potential for collusion.

\subsection{System Liveness}
To ensure system liveness and resilience against attacks such as reshipping and settlement inconsistencies, the implementation of a robust recovery and dispute resolution protocol is essential in the \verb|SecurePay| framework. 

\textbf{Issue 1. Reshipping Attack:} A malicious buyer can exploit the delivery confirmation process within the smart escrow contract, denying receipt of the product to force the merchant to reship it. 

\textbf{Attack Details:} During a reshipping attack, a malicious buyer may fail to confirm product delivery, recording their decision as 'Null' with their digital signature within the blockchain. This lack of confirmation prevents the smart contract from advancing to the settlement phase, which requires the buyer's confirmation of receipt.

\textbf{Defense Protocol:} To counter this, when a delivery confirmation is recorded as 'Null,' the smart contract initiates a dispute resolution phase. In this phase, both the buyer and the courier are required to deposit an amount equal to the product's price, ensuring that one of them acts in good faith regarding the product's delivery. It is assumed that couriers have verifiable proof of delivery, which they submit through the E-commerce platform. The platform then takes on the role of resolving the dispute off-chain and deciding the outcome. The party deemed honest is refunded their deposit, plus they receive half of the dishonest party's deposit. The remaining portion of the deposit is allocated to the E-commerce platform's wallet. This approach not only makes delivery-related attacks financially unfeasible but also provides a clear path to proceed smart contracts.

\textbf{Issue 2. Settlement Inconsistency:}
Banks are mandated to transfer funds to the merchant's wallet based on the ledger update on the blockchain. However, manual errors such as missing certain transactions or inputting incorrect numbers can result in discrepancies between the on-chain and off-chain records, leading to settlement inconsistencies.

\textbf{Resolve mechanism:} To address this vulnerability, we leverage the traceability and programmability of CBDCs (see Appendix A) to automate settlement processes based on blockchain data and create verifiable evidence. In case of a suspected settlement discrepancy, merchants can challenge unsettled transactions by submitting the transaction ID and requesting proof of settlement within their CBDC accounts. Banks must then verify the settlement by tracing the escrow funds linked to the transaction and confirming the correct settlement into the merchant's account. Failure to provide this proof results in the bank compensating the merchant financially. This approach maintains settlement integrity by utilizing the transparency and traceability of CBDC transactions.

\subsection{User Privacy}
We leverage the privacy-preserving capabilities of CBDCs \cite{bis,hamilton, benigno} to safeguard user privacy during transactions. In these systems, only the central bank has access to the real identity of the account holder, limiting E-commerce platforms' access to authenticated buyer information. Banks alone are responsible for verifying payments off-chain, ensuring that wallet balances and transaction details remain undisclosed to unauthorized parties. Furthermore, we implement the certificate management feature in a permissioned blockchain to regulate authorized access to transaction information \cite{hyperledger}. Only approved mediators and merchants are granted the ability to validate a buyer's transaction proposal, ensuring confidential transaction data is protected from unauthorized access by other participants.

\section{implementation and evaluation}

We adopt Hyperledger Fabric (Fabric) \cite{hyperledger}, a prominent permissioned blockchain platform, as the foundation for our smart escrow service system. Hyperledger Fabric is an open-source framework designed to support diverse business applications through the deployment of smart contracts, known as Chaincodes \cite{hyperledger}. Unlike Ethereum, which requires domain-specific languages for smart contract development, Fabric allows distributed applications to be written in multiple programming languages, offering greater flexibility. The consensus protocol in Hyperledger Fabric is modular, catering to various trust models and use cases without relying on a native cryptocurrency \cite{hyperledger}. This modularity enables the customization of distributed applications based on specific business requirements. Within Fabric, organizations establish permissioned blockchains through designated channels, each maintaining its own ledger and governance mechanisms \cite{hyperledger}. Secure identity management is enforced via a Certificate Authority (CA) and Membership Service Provider (MSP) \cite{hyperledger}, ensuring that only authorized entities—such as the bank, and platform—can collaborate within the network to maintain, process, and cross-validate wallet balances for E-commerce transactions, protecting transaction privacy.

We integrate the OpenCBDC payment system \cite{hamilton}, developed under Project Hamilton, as our CBDC off-chain payment processing system. While OpenCBDC offers two architectures for transaction processing - the Atomizer architecture and the Two-phase commit (2PC) architecture \cite{2pcc} - we opt for the 2PC architecture in our implementation. This choice is driven by two key factors: firstly, the ordering sequence is already established through consensus on Hyperledger Fabric, eliminating the need for the total ordering provided by the Atomizer structure; secondly, our system's performance is primarily constrained by Hyperledger Fabric, rendering the choice between Atomizer and 2PC architectures inconsequential for our performance testing. To bridge the on-chain chaincode with the off-chain OpenCBDC system, we implement approximately 300 lines of Go code using the Hyperledger Fabric SDK \cite{sdk}, ensuring seamless integration and execution of settlements.

For benchmarking and testing, we deployed a Hyperledger Fabric network (version 2.2.0) consisting of 1 orderer and 2 peers to implement an E-commerce smart contract developed in Golang (version 1.0.1). The deployment environment includes a laptop equipped with an AMD Ryzen 7 5800H processor (3.20 GHz) and 16GB of RAM, alongside servers running Ubuntu 20.04 hosted on VMware Workstation, each configured with 4-core CPUs and 11.3GB of RAM. Our source code is publicly available at \cite{cbdc}. To evaluate the performance of our system, we utilize Hyperledger Caliper \cite{caliper}, an open-source benchmarking tool designed to assess the performance of various blockchain platforms, including Hyperledger Fabric, Ethereum, and Corda. We compare the \verb|SecurePay| framework with existing solutions \cite{fy, goldfeder,liu} to fully reveal the necessity and superiority of our framework.

In the next sections, we will assess our system's performance using transaction throughput (TPS) and latency. The evaluation includes three areas: normal conditions, potential attacks including reshipping attack and settlement inconsistency, and optimization strategies to improve TPS and reduce latency.

\begin{figure}[htbp]
  \centering
  \includegraphics[width=\linewidth]{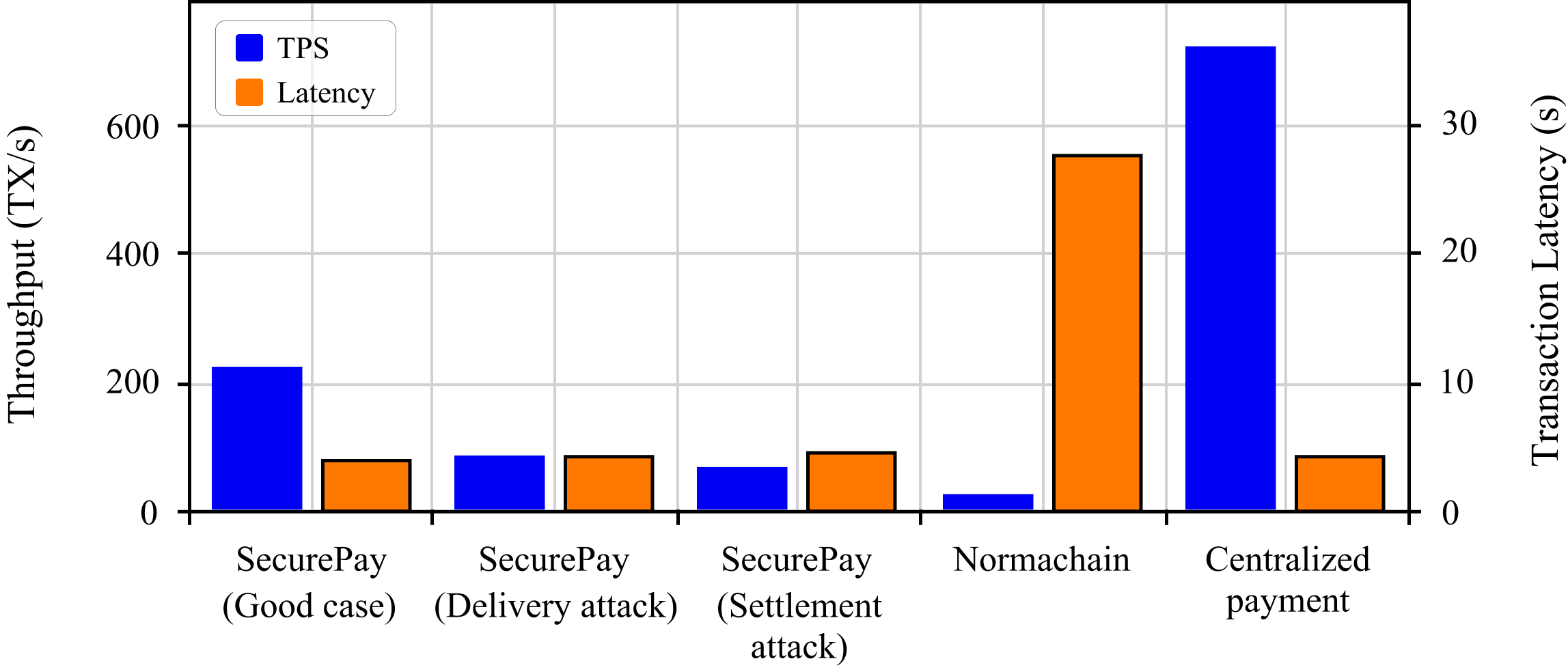}
  \caption{SecurePay system performance under normal conditions and the performance under potential attacks. The system performance primarily based on transaction throughput (TPS) and average latency.}
  \label{sys_per}
\end{figure}

\subsection{Good Case Performance}
Figure \ref{sys_per} presents a comparison of system performance with existing payment systems \cite{liu, fy, goldfeder}. We have fully implemented the \verb|SecurePay| workflow (Figure \ref{sysover}) and tested transaction throughput (TPS) and latency. To accurately simulate an E-commerce platform for our tests, we developed a class specifically designed to perform key database operations common in E-commerce platforms. This class, though streamlined, encapsulates essential functions such as creating orders, managing inventory, and processing payments, mirroring the core functionalities without the complexities of a full-scale E-commerce database system. Furthermore, we replicated the smart contract service and settlement mechanisms of Normachain \cite{liu} and deployed them on the Ethereum blockchain. In an effort to replicate a centralized payment system, we utilized our simulated online E-commerce platform database in conjunction with OpenCBDC \cite{hamilton} to act as a surrogate for the Real-Time Gross Settlement (RTGS) system for processing monetary transactions.

\verb|SecurePay| achieves a transaction throughput of 256.4 transactions per second (TPS) with an average latency of 4.29 seconds for processing E-commerce transactions. Our system demonstrates high efficiency in handling transaction data, resulting in the lowest latency compared to other payment systems. Additionally, our system maintains a competitive transaction throughput without significant compromise on performance when compared to centralized payment systems like Real Time Gross Settlement (RTGS) systems \cite{fastpay}, which can reach 774.2 TPS.

\subsection{Performance under Attacks}

\textbf{Reshipping Attack.} We simulate the empirical reshipping attack by the buyer not submitting the product delivery and triggering the dispute phase of the smart contract. We simulate the process of the buyer and courier depositing to the dispute wallet. We assume the E-commerce platform has finished solving the dispute off-chain and uploads the winner from their digital signature on-chain to proceed with smart contracts. The winning party will receive compensation from the dispute wallet, and the smart contract will proceed to product confirmation for settlement. In this case, \verb|SecurePay| will only be able to process transactions at a TPS of 86.4 (TX/s) and increase the average latency to 4.81 seconds.

\begin{figure}[htbp]
  \centering
  \includegraphics[width=\linewidth]{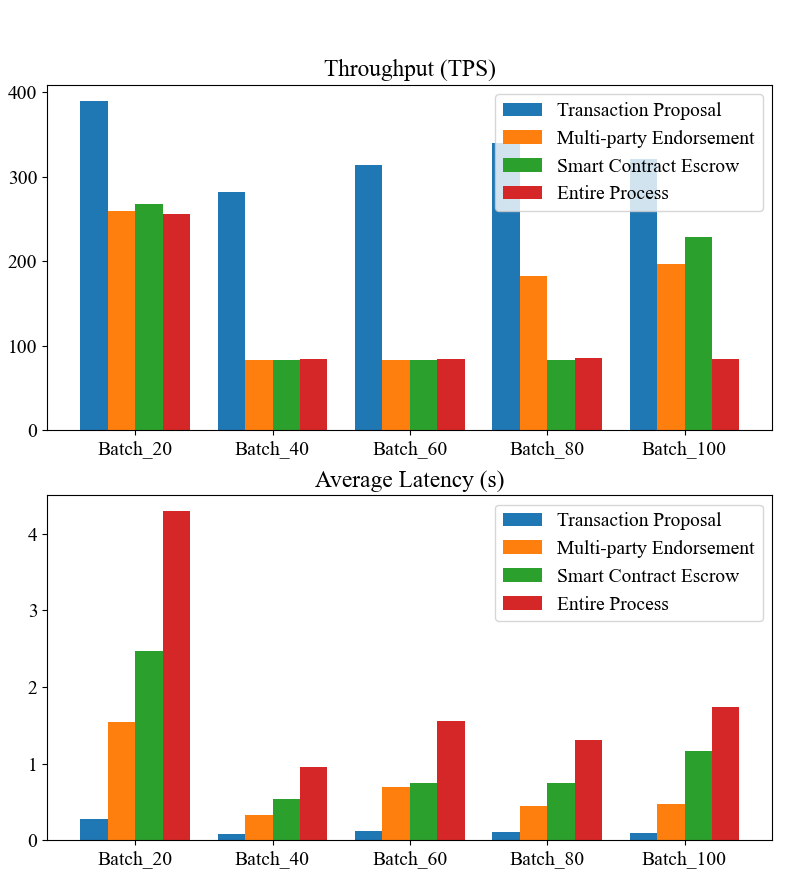}
  \caption{Throughput (TPS) and average latency of the SecurePay payment system across different batch sizes, separated from the individual components of the system workflow (transaction proposal, multi-party endorsement, and smart contract escrow). }
  \label{sys_opt}
\end{figure}

\textbf{Settlement Inconsistency.} We simulate the process of requiring the bank to generate proof for each settlement from CBDC transactions to avoid settlement inconsistency. OpenCBDC provides \textit{importstring}, a cryptographic proof for payment for recipients to redeem the fund from the CBDC wallet \cite{hamilton}. Thus, we require the bank to upload \textit{importstring} for every transaction, which drives down the TPS to 81.4 (TX/s) and the average latency to 5.07 seconds.

\subsection{System Optimization}
Our system is segmented into three key stages as outlined in the system workflow (Figure \ref{sysover}): transaction proposal, multi-party endorsement, and smart contract escrow. To assess performance, we evaluated each stage independently by varying the batch size—the maximum number of transactions that a single block can contain \cite{hyperledger}.

As demonstrated in Figure \ref{sys_opt}, increasing the batch size leads to a decrease in transaction throughput, with a notable reduction in average latency between a batch size of 20 and 40, followed by a slight increase in latency as the batch size continues to grow. Our system defaults to a batch size of 20 to optimize smart contract processing. Although a larger batch size can improve latency, \verb|SecurePay| already outperforms existing payment systems in terms of transaction latency \cite{liu, fy, goldfeder}. Therefore, our primary focus is to maintain a transaction throughput comparable to that of centralized payment systems. At a batch size of 20, our system achieves its optimal transaction throughput of 256.4 transactions per second, which aligns with our strategic objectives.

A major bottleneck in transaction throughput is observed during the multi-party endorsement phase, which results in the lowest TPS (Figure \ref{sys_opt}). In this phase, all involved parties must broadcast their decisions on the buyer's proposal and wait for the blockchain nodes to update the batch of transactions. Future enhancements will focus on optimizing smart contract logic, including implementing batch signatures and batch payment information updates to streamline this process.

\section{Conclusion}
In this paper, we introduce \verb|SecurePay|, a payment processing system to ensure security goals including fund security, information security while maintaining fast transaction processing. We utilize a multi-party endorsement algorithm to preserve information integrity, the programmability of CBDCs to secure funds, and privacy-preserving features inherent in CBDCs to protect user privacy. The hybrid on-chain/off-chain execution model not only safeguards user wallet information but also delivers superior system performance compared to existing solutions. We believe that advancements in smart contract technology will revolutionize payment processing within the platform economy by facilitating efficient multi-party collaboration, increasing the level of automation, ensuring transaction accuracy, and maintaining user privacy.

\section{Acknowledgment}
The work of Songze Li is in part supported by the Fundamental Research Funds for the Central Universities (Grant No. 2242025K30025). Xuechao Wang is supported by the Guangzhou-HKUST(GZ) Joint Funding Program (No. 2024A03J0630 and No. 2025A03J3882), the Guangzhou Municipal Science and Technology Project (No. 2025A04J4168), and a gift from Stellar Development Foundation.

\appendix
\section{Central Bank Ditital Currency}

CBDC represents a digital currency issued and managed by central banks, offering a cash-like peer-to-peer payment system. Designed to coexist with physical cash, it is guaranteed by the central bank for redemption and is backed by a 1-to-1 reserve in fiat currency \cite{auer}. CBDC presents several advantages over other means of payments (Figure \ref{comp}):

\begin{figure}[h]
    \centering
    \includegraphics[width=0.48\textwidth]{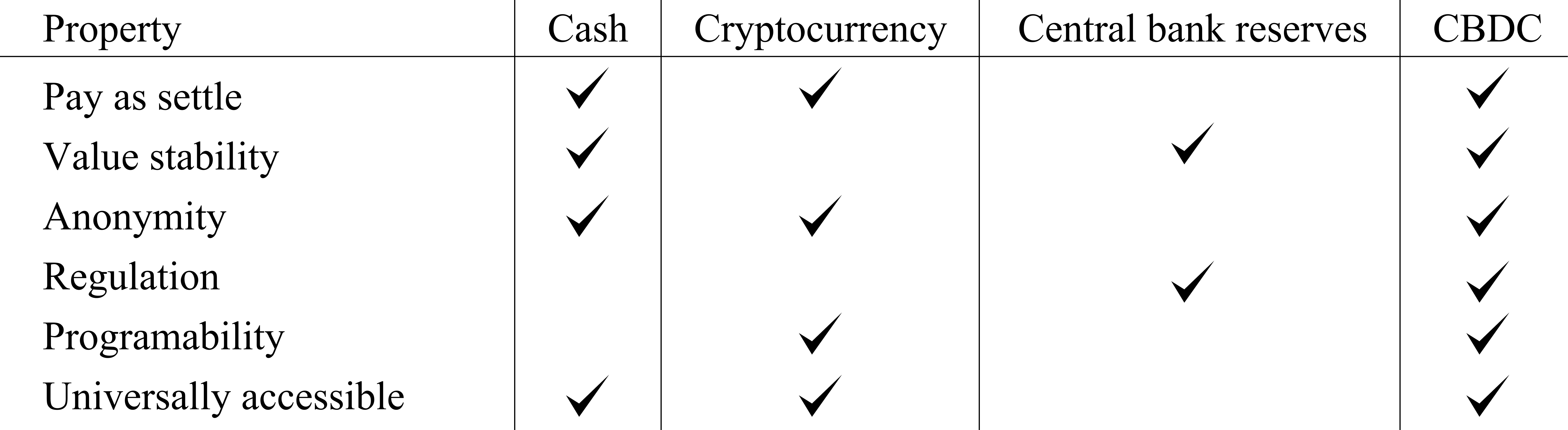}
    \caption{Table summarizing the properties of various monetary instruments \cite{hamilton, bis}.}
    \label{comp}
\end{figure}

\textbf{Instant settlement}. Unlike platform settlements, which rely on an account system and financial intermediaries, CBDC enables instant settlement for any transaction amount as a token-form digital currency similar to cryptocurrency and cash \cite{auer,fy}.

\textbf{Value stability}. Backed by a 1-to-1 reserve in fiat currency, CBDC offers stable value as a means of payment, unlike cryptocurrencies, which are subject to high price volatility due to market events \cite{kumar, keister}. 

\textbf{Anonymity}. CBDC protects consumers' privacy by enabling anonymous payments, in contrast to central bank reserves and cryptocurrencies, which offer varying degrees of privacy and anonymity \cite{bis, benigno}. 

\textbf{Regulatory environment}. CBDC can incorporate compliance with existing regulations, such as AML/CFT requirements and consumer protection laws, into its design, a feature challenging for decentralized and anonymous cryptocurrencies \cite{imf}.

\textbf{Programability}. Both CBDC and cryptocurrency utilize smart contracts for programmable account settlement, limiting fund settlement to pre-determined wallets under supervision, thus preventing misuse \cite{imf}.

\section{Comparison of existing payment systems}\label{compa}

This section presents a comparative analysis of existing payment systems, evaluating them across four critical dimensions: fund security, information security, stability of currency value as a medium of exchange, and system performance (specifically transaction throughput and latency). The assessment employs a tripartite scale (low-medium-high) for each dimension, facilitating a comprehensive yet accessible comparison. To enhance the visualization of these multifaceted comparisons, the results are presented in a radar chart (Figure \ref{compar}). This graphical representation allows for an intuitive understanding of the relative strengths and weaknesses of each payment system across the evaluated criteria, enabling a holistic view of their performance in the context of modern financial ecosystems.

\subsection{Centralized payment system \cite{fy}}

The centralized payment system demonstrates varying levels of efficacy across the evaluated dimensions. In terms of fund security, it exhibits medium performance due to the involvement of regulatory institutions like banks in managing escrow funds, effectively preventing unauthorized use \cite{fastpay}. However, due to the escrow and settlement process is not transparent to both buyers and merchants from their specific transaction, it can not present full security of fund. The information security presents a notable vulnerability. While transaction details are recorded, they are primarily managed by E-commerce platforms, introducing the risk of manipulation, such as unauthorized changes to recipient addresses, potentially exposing users to significant financial losses. 

The system's strength lies in its ability to maintain currency stability through the use of fiat currency facilitated by the Real-time Gross Settlement System (RTGS), an inter-bank transaction network that operates on central bank reserves \cite{fastpay}. This setup ensures a high level of stability in the value of the traded currency. Furthermore, the system demonstrates exceptional performance, with the RTGS supporting a retail-level settlement network capable of processing up to 80,000 transactions per second and achieving payment finality with latency under 200ms \cite{fastpay}.

In terms of privacy protection, the system offers moderate safeguards through the user data protection policy enforced by the E-commerce platform. However, a potential vulnerability arises from the fact that transaction information processed by the E-commerce platform is shared with various stakeholders such as banks and courier companies without adequate supervision and technical protection. This lack of oversight and protection could lead to unauthorized access to transaction details, posing a risk to user privacy and data security.

\subsection{Encrypt-and-swap \cite{goldfeder}}
Goldfeder et al. \cite{goldfeder} introduce a novel decentralized escrow protocol utilizing Bitcoin, which employs an "encrypt-and-swap" method for settling escrow transactions. This protocol initiates when both the buyer and merchant generate a shared 2-of-2 ECDSA key. The security mechanism involves each party encrypting their portion of the key using the mediator's public key and then exchanging these for verification purposes. The mediator's role is critical as they validate escrow payments and determine which party is eligible to redeem the funds. Redemption occurs when the buyer confirms receipt of goods, prompting the mediator to forward the key shares to the merchant.

While this protocol robustly secures the escrow funds—restricting the mediator from redeeming the funds themselves—it does not safeguard transaction information adequately. The process of mediator appointment by the buyer and seller introduces potential risks, such as the possibility of collusion with one of the parties to endorse false transaction information, thereby enabling a malicious party to unjustly redeem escrow funds \cite{goldfeder}.

The system utilizes Bitcoin as the medium of exchange, known for its price volatility, which may not be suitable for daily E-commerce transactions due to significant value fluctuations. Additionally, the system's performance is notably slow, with transaction completion experiencing substantial latency, averaging around 140 minutes per transaction \cite{yang}, limiting its practicality for frequent or time-sensitive transactions.

In terms of system privacy, the protocol offers a moderate level of protection as blockchain nodes without authorization cannot access escrow payment information due to the encryption method used in Bitcoin \cite{yang}. However, transaction amounts can be inferred, and user wallet balances traced through specific transactions, posing potential privacy vulnerabilities \cite{endurthi}.

\subsection{Normachain \cite{liu}}

Liu et al. \cite{liu} present Normachain, a three-layer blockchain architecture enhancing decentralized escrow protocols. The transaction layer initiates the process, allowing buyers and merchants to generate contracts with critical details like pricing, transaction IDs, and wallet addresses, ensuring information security through tamper-proof recordings.  The consortium approval layer, managed by banks, verifies and settles transactions via distributed consensus, while the supervision layer employs searchable encryption to identify potentially illegal transactions. This comprehensive structure streamlines the recording, settling, and supervision of transactions, significantly improving the efficiency, security, and regulatory compliance of decentralized escrow processes.

Despite these innovative features, Normachain lacks a robust escrow service that can securely lock money transfers in a digital wallet inaccessible to any intermediaries, including banks. This limitation implies that while the banks involved have no direct incentive to compromise user trust, the system only provides a medium level of fund security. Moreover, Normachain has introduced its cryptocurrency, NorMaCoin, as the trading currency. This choice subjects the system to the typical volatility associated with cryptocurrencies, which is largely dependent on public belief and market dynamics. In terms of system performance, Normachain operates at a medium level, attributed to its design of a three-layer network that segregates transaction information from payment processing across different chains.

One notable concern with Normachain is the lack of emphasis on protecting user data privacy. Since all user transactions are publicly recorded on the transparent transaction layer, the public blockchain, through the execution of smart contracts using the Normachain template, there is a significant risk to user data privacy. The exposure of transaction details on a public blockchain can compromise user anonymity and confidentiality, potentially leading to privacy breaches and security vulnerabilities.
\end{document}